\documentclass[a4paper,12pt]{article}
\usepackage{rotating}
\usepackage{graphicx}
\usepackage{wrapfig}
\usepackage{braket}
\usepackage{amssymb,amsmath,bm}
\usepackage{setspace,comment,lipsum} 
\usepackage[colorlinks, citecolor=blue,anchorcolor=red,menucolor=red, linkcolor=red,filecolor=red,runcolor=red,urlcolor=blue,frenchlinks=red]{hyperref}
\usepackage{authblk}

\def\be{\begin{equation}}
\def\ee{\end{equation}}
\def\bdm{\begin{displaymath}}
\def\edm{\end{displaymath}}
\def\L{\mathcal{L}}

\def\d{\partial}

\def\dsds{D_s^*\bar{D}_s^*}
\def\vpsif0{J/\psi f_0(980)}
\def\vpsi2d{\psi(4160)}

\providecommand{\keywords}[1]{\textbf{\textit{Keywords --}} #1}

\title{On the origin of the $Y(4260)$}
\date{}
\author[1]{S. Coito \footnote{Electronic address: scoito@ujk.edu.pl}}
\author[1,2]{F. Giacosa \footnote{Electronic address: fgiacosa@ujk.edu.pl}}
\affil[1]{\it\small Institute of Physics, Jan Kochanowski University, 25-406 Kielce, Poland}
\affil[2]{\it\small Institut f\"ur Theoretische Physik, Johann Wolfgang Goethe-
Universit\"at, 60438 Frankfurt am Main, Germany}

\begin{document}
\maketitle
\onehalfspacing
\begin{abstract}
We study the relation between the $\psi(4160)$ and the $Y(4260)$ within an unitarized effective Lagrangian approach. The $Y(4260)$ arises as a manifestation of the $\psi(4160)$, when a loop-driven decay of the type $\psi(4160)\to D_s^*\bar{D}_s^*\to J/\psi f_0(980)$ is enhanced by the proximity of the pole, corresponding to the $\psi(4160)$, to the {\it almost} closed $D_s^*\bar{D}_s^*$ decay channel. Other $f_0$ resonances that may add a non-negligible contribution, by the same mechanism, are not included for simplicity, but they are not expected to change the main conclusion. Within this picture, the $Y(4260)$ is not, therefore, an independent resonance, but rather a variation of the $\psi(4160)$, which also explains why it is not seen in OZI-allowed decay channels in the experiment. 
\end{abstract}

\\

\maketitle



\section{Introduction}

The $Y$ ``states'' are enhancements in the vector charmonium mass distribution, that: i) are seen in the suppressed modes only (viz.~Okubo-Zweig-Iizuka (OZI)-suppressed), whereas the regular $\psi$ are not; ii) are not seen in the dominant modes with open-charm (viz.~OZI-allowed), whereas the $\psi$ are; iii) their mass is very close to, yet not coincident with, the mass of the $\psi$. Such characteristics are intriguing, as they point out nonperturbative phenomena outside of the quark model, that cannot accommodate so many states with the same quantum numbers and similar mass. The $\psi$ excitations, up to about 4.5 GeV, have been known from general fits to $R$ data \cite{plb660p315,prd72p017501}. The $Y$ signals have shown in modes such as $J/\psi\pi^+\pi^-$ \cite{prl118p092001,prl110p252002}, $\psi(2S)\pi^+\pi^-$ \cite{prd91p112007}, $h_c\pi^+\pi^-$ \cite{prl118p092002}, and $\omega\chi_{c0}$ \cite{prd99p091103}. To each peak, in each one of these channels, a different $Y$ has been assigned \cite{pdg}. Such separation is made due to the fact that, in different modes, Breit-Wigner fits of the the signals lead to different parameters of mass and width, yet it is very plausible that some of these peaks actually correspond to the same resonance. Different manifestations of the same pole in different channels is a know phenomenon, since the interference with the background will be different. As an example, the cross sections of the $\psi$ excitations in channels $D\bar{D}$, $D\bar{D}^*+c.c.$, and $D^*\bar{D}^*$ \cite{prd77p011103,prd79p092001}, present different line-shapes. 

There is a clear enhancement in the $J/\psi\pi^+\pi^-$ invariant mass distribution, with mass between 4.22-4.28 GeV and width between 40-140 MeV, namely the $Y(4260)$ \cite{prl118p092001,prl110p252002}. Its average mass and width is $4230\pm 8$ MeV and $55\pm 19$ MeV, in the latest version of PDG \cite{pdg}. There have been indications of a similar enhancement in the mode $J/\psi K^+K^-$ \cite{prd97p071101}. On its hand, the mass of the $\psi(4160)$ has average mass and width $4191\pm 5$ GeV and $70\pm 10$ MeV, correspondingly, thus only about 40 MeV below the mass of the $Y(4260)$. The branching fraction of the $\psi(4160)$ to $J/\psi\pi^+\pi^-$ is no more than 0.3$\%$, in spite of the large phase-space available, so it is practically not seen in this channel. Also intriguing, is the recent observation of a clear signal in $\pi^+D^0D^{*-}$ channel with mass and width at about 4.23 GeV and 77 MeV, respectively, but with no traces of the $\psi(4160)$ \cite{prl122p102002} (it is however not clear if the resonance found in this work shall be assigned to the $Y(4260)$ or to another novel state, such as the $Y(4220)$). 

The nature of the $Y(4260)$ has been explored in different approaches. In Refs.~\cite{0904.4351,prd79p111501,prl105p102001}, the $Y(4260)$ enhancement is seen as the result of interference phenomena between the channels $\dsds$ and $J/\psi f_0(980)$, thus it is not regarded as a true resonance. In Refs.~\cite{prd83p054021,prd93p014011}, a similar nonresonant hypothesis is analyzed, but through an interference between the vectors $\psi(4160)$ and $\psi(4415)$, including the $D\bar{D}$, $D\bar{D}^*+c.c.$, and $D^*\bar{D}^*$ loops, successfully reproducing the line-shape of the $Y(4260)$. A different approach considers a hadrocharmonium, i.e., a charmonium embedded in a sea of light quarks, where the $Y(4260)$ is a mixture between $^3S_1$ and $^1P_1$ charmonium states, with the $Y(4360)$ as its pair \cite{mpla29p1450060}. Other approaches consider the $Y(4260)$ to be a resonance of ``molecular'' type, where a $\bar{c}c$ core is coupled mainly to the $D\bar{D}_1+c.c.$ channel, see Refs.~\cite{prd96p114022,prd89p034018,epja52p310,plb768p52,prd94p054035,prd90p074039}, thus with an important decay to channel $Z_c^\pm(3900)\pi^\mp\to J/\psi \pi^+\pi^-$.  Dynamical generation is also studied in Ref.~\cite{prd80p094012}, using both the $J/\psi \pi^+\pi^-$ and $J/\psi K^+K^-$ systems, with the emergence of a resonance around 4.15 GeV, rather closer to the $\psi(4160)$ state. Tetraquark models may be found in Refs.~\cite{prd72p031502,epjc78p29}. Reviews on the $Y$ resonances are found in \cite{pr668p1,rmp90p015004,1907.07583}.

In this work, we present a novel result in which the $Y(4260)$ and the $\psi(4160)$ correspond to the same resonance, i.e., to the same pole, but with two different peaks in different channels. The underlying mechanism for the generation of the $Y(4260)$, within this study, is the decay chain $\vpsi2d\to\dsds\to J/\psi f_0(980)\to J/\psi \pi^+\pi^-$. We stress, however, that other channels can lead to the final state $J/\psi \pi^+\pi^-$, namely all those involving scalar mesons, such as the $f_{0}(500),$ $f_{0}(1370),$ $f_{0}(1500),$ and $f_{0}(1710)$, yet we do not include them for simplicity (the $f_{0}(980)$ is expected to give the biggest contribution among the $f_0$ family, for reasons that we shall explain in the next sections). This process is enhanced by two factors: (i) the mass $\psi (4160)$ lies just below
the $D_{s}^{\ast }\bar{D}_{s}^{\ast }$ threshold and (ii) the contribution
of the $D_{s}^{\ast }\bar{D}_{s}^{\ast }$ loop is enhanced just at its
threshold. Both properties, when simultaneously realized, can {\it shift} the peak produced by the pole of the $\psi(4160)$ to a value close to $4.23$ GeV, in the $J/\psi \pi ^{+}\pi ^{-}$ channel. Although other $f_0$ mesons, excluding the $f_0(980)$, may play a non-negligible role, the main point is that the contribution of the $D_{s}^{\ast }\bar{D}_{s}^{\ast }$ loop is enhanced at its threshold (this is a peculiarity of the real part of the loop), thus all $f_0$ should give rise to a similar peak position in the final $J/\psi \pi ^{+}\pi^{-}$ channel, moved from the original $\psi(4160)$ position to a value of about $4.23$ GeV. Such result opens the possibility that the excess of vectorial resonances seen in the experiment might be largely fictitious, while it also helps to understand the unquenching of the vector charmonia. Our main result shows the possibility of generating an amplitude peak with a ``shifted'' mass, that manifests in a certain channel. Preliminary studies of the current work may be found in Refs.~\cite{1810.03532,1811.08183}. This idea is also aligned with the phenomenology of a recent analysis from JPAC group, where it was found that, subjacent to the $\pi_1(1400)$ and $\pi_1(1600)$ resonances, there is only one pole \cite{prl122p042002}.

This paper is organized as follows: In Sec.~\ref{model} we describe briefly the unitary effective Lagrangian model we employ, that includes loop meson-meson loops in the propagator function. In Sec.~\ref{description} we employ the model to the description of the vector $\psi(4160)$. In Sec.~\ref{interference} the decay of the $\psi(4160)$ to channel $\vpsif0$ is explored, either in case of the direct decay, Sec.~\ref{direct}, and via $\dsds$ loops, in Sec.~\ref{inter}, which is the main result of our paper. In Sec.~\ref{conclusion} we draw the conclusion.


\section{\label{model}The model}

\begin{figure}
\begin{center}
\hspace*{-0.6cm}\resizebox{!}{75pt}{\includegraphics{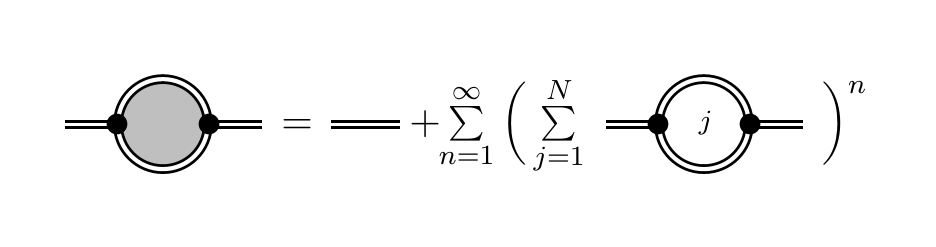}}
\end{center}
\caption{\label{sum}Scheme of the propagator with one seed and $N$ channels. See text for details.}
\end{figure}

The unitary effective Lagrangian model that we employ here is described in detail in Refs.~\cite{prd93p014002}, \cite{npa931p38}, \cite{epjc79p98}, and \cite{ijmpa34p1950173} respectively to systems $a_0(980)$ and $K_0^*$, $\psi(3770)$, $\psi(4040)$, and $X(3872)$. A similar formalism is also found in Ref.~\cite{1901.09862}. A single vector meson, e.g.~produced in an annihilation experiment, is propagating in momentum space. Yet, rather than a simple $\bar{q}q$ system, the vector meson is dressed with OZI-allowed meson-meson loops, according to the the scheme in Fig.~\ref{sum}. Each loop $j$ is a different meson-meson channel in a total of $N$ channels (cf.~Table \ref{channels}). The sum over $n$ is the equivalent to the Born series within scattering theory, thus it obeys a geometric progression. The first term on the right side of the figure represents the propagation of the undressed $\bar{q}q$ seed state with mass $m_0$. The scalar part of the full propagator of the dressed seed state is written as 
\be
\label{prop}
\Delta_\psi(s)=\frac{1}{s-m_0^{2}+\sum_j^N\Pi_j(s)}\ ,
\ee
where $s$ is the invariant mass squared of the vector meson, and $\Pi_j(s)$ is the loop function of channel $j$, that is given by
\be
\label{loopf}
\Pi_j(s)=\Omega_j(s) + i\sqrt{s}\Gamma_j(s),\ \ \Omega,\ \Gamma \in \Re\ ,
\ee
where the real part is given by the dispersion relations
\be
\label{omega}
\Omega_j(s)=\frac{PP}{\pi}\int_{s_{thj}}^{\infty}\frac{\sqrt{s'}\Gamma_j(s^{\prime})}%
{s^{\prime}-s}\ \mathrm{d}s^{\prime},
\ee
with $th$ as the abbreviation for threshold. The imaginary part in Eq.~\eqref{loopf} is given by 
\be
\label{gamma}
\Gamma_{j}(s)=\frac{k_j(s)}{8\pi s}
|\mathcal{M}_{\psi\to(m_1m_2)_j}|^{2}\ , \mathrm{with}
\ee
\be
\label{amp}
|\mathcal{M}_{\psi\to(m_1m_2)_j}|^{2}=\mathcal{V}_{\psi\to(m_1m_2)_j}(s)\ f_{\Lambda}^{2}(\vec{k}_j^2)\ .
\ee
In Eq.~\eqref{gamma}, $k_j(s)\equiv k(s,(m_1,m_2)_j)$ is the relativistic center-of-mass momentum of channel $j$, depending on the masses $m_1$ and $m_2$ of the meson-meson pair. In Eq.~\eqref{amp} $\mathcal{V}$ are the 3-vertex amplitudes, represented by black circles in Fig.~\ref{sum}, which are computed using the Feynman rules, given the interaction Lagrangians. The function $f$ is a vertex form-factor that depends on a cutoff parameter $\Lambda$ and on the momentum, and it is here defined by an exponential function as
\be
\label{ff}
f_{\Lambda}(\vec{k}_j^2)=e^{-\vec{k}_j^2/\Lambda^2}\ .
\ee
We note that $f$ is only a partial form factor, since the full vertex amplitude in Eq.~\eqref{amp} is given by the product of $\mathcal{V}$ with $f^2$. Therefore, it cannot be directly compared to form-factors that represent the whole charge distribution of a certain composite particle, such as the Sachs electric, magnetic, or quadrupole form factors, that have been used in lattice QCD calculations, as for instance in Ref.~\cite{plb719p103}. For a detailed treatment of the form-factor, see Ref.~\cite{npa931p38} and refs.~therein.

The full spectral function is given by 
\be
\label{sfi}
d_{\psi}(\sqrt{s}) =-\frac{2\sqrt{s}}{\pi}\mathrm{Im}\ \Delta_\psi(s)\ ,
\ee
which due to the unitarity comes automatically normalized to 1, i.e.~$\int_0^\infty \mathrm{d}\sqrt{s}\ d_\psi(\sqrt{s})=1$. Explicitly, $d_\psi(\sqrt{s})$ reads:
\be
\label{sf}
d_{\psi}(\sqrt{s})=\frac{2s}{\pi}\frac{\sum_j^N\Gamma_j(s)}{\Big[s-m_0^{2}+\sum_j^N\Omega_j(s)\Big]^{2}+\Big[\sqrt{s}\sum_j^N\Gamma_j(s)\Big]^2}\ ,
\ee
whereas each partial spectral function is given by 
\be
\label{sfp}
d_{\psi\to(m_1m_2)_l}(\sqrt{s})=\frac{2s}{\pi}\frac{\Gamma_l(s)}{\Big[s-m_0^{2}+\sum_j^N\Omega_j(s)\Big]^{2}+\Big[\sqrt{s}\sum_j^N\Gamma_j(s)\Big]^2}\ ,
\ee
i.e.~$d_{\psi}(\sqrt{s})=\sum_j^N d_{\psi\to(m_1m_2)_j}(\sqrt{s})$. It can be noticed that, since the denominator is the same for each partial spectral function, the line-shape in each channel will vary solely through the shape of the decay function $\Gamma_j(s)$. Typically, but not always, the peak is centered is at about $m_\psi^{peak}$ for each partial spectral function.\\

The poles are computed through the analytic continuation to the complex plan of $s$, i.e. $s\to z^2$, by solving
\be
\label{poles}
z^2-m_0^{2}+\sum_j^N\Pi_j(z^2)=0,\ z\in\mathbb{C}\ ,
\ee 
where the function $\Pi_{j}(s=z^{2})$, in its first Riemann
sheet, reads:
\begin{equation}
\Pi_{j}(s)=\frac{1}{\pi}\int_{s_{th,j}}^{\infty}\frac{\sqrt{s^{\prime}}%
\Gamma_{j}(s^{\prime})}{s^{\prime}-s}\ \mathrm{d}s^{\prime},\text{ }s\in\mathbb{C}
\text{ .}%
\end{equation}
One may note that $\Pi_{j}(s)$ is regular everywhere on the complex $s$-plane:
apart from the cut from $s_{th,j}$ to $\infty$ on the real axis, there is no
pole or other singularity. Note, this is true for any chosen form factor,
including the exponential one introduced previously. In particular, $\Pi
_{j}(s\rightarrow\infty)\rightarrow0$ in all directions. In this context, it
is important to recall that, in the 1st Riemann Sheet, the function $\Pi
_{j}(s)$ is an utterly different complex function than $f_{\Lambda}^{2}%
(k_{j}^{2})\propto e^{-2k_{j}^{2}/\Lambda^{2}}$. Namely, $\operatorname{Im}%
\Pi_{j}(s)=\sqrt{s}\Gamma_{j}(s)$ is solely valid for $s$ being real. This
fact is also clear by noticing that, while $\Pi_{j}(s\rightarrow
\infty)\rightarrow0$ in any direction, this is not the case for the form
factor, which, in the exponential case, has an essential singularity for
$s\rightarrow\infty$.

On the second Riemann sheet, for the $j$-th channel, the loop function reads:
\begin{equation}
\Pi_{j,II}(s)=\Pi_{j}(s)+2i\sqrt{s}\Gamma_{j}(s)\text{ .}%
\end{equation}
There is in this respect a simple subtle point: the complex function $\sqrt
{s}\Gamma_{j}(s)\propto\sqrt{s}\sqrt{s-s_{th,j}}$ has two cuts, from $-\infty$
to $0$ and from $s_{th,j}$ to +$\infty.$ When $\Pi_{j,II}(s)$ is taken in the
second Riemann sheet, one should take $\sqrt{s}\Gamma_{j}(s)$ on its II
Riemann sheet as well. As a consequence, for $s=x^{2}+i\varepsilon,$
$\operatorname{Im}\Pi_{j}(x^{2})=\sqrt{x^{2}}\Gamma_{j}(x^{2})>0,$ while
$\operatorname{Im}\Pi_{j,II}(x^{2})=-\sqrt{x^{2}}\Gamma_{j}(x^{2})<0.$

Next, the full loop function in the first Riemann sheet reads $\Pi
(s)=\sum_{j=1}^{N}\Pi_{j}(s)$, where it is useful to use the ordering
$s_{th,i}<s_{th,j}$ for $i<j.$ For each term $\Pi_{j}(s)$ one can take the I
or the II Riemann sheet, for a total of $2^{N}$ possibilities. Most of the
Riemann sheets, however, are not useful for our analysis. The interesting
poles (close to the real axis) for a certain energy interval of
\ $\operatorname{Re}[s]$ are typically obtained by considering the second
Riemann sheet for all the channels $\Pi_{j}(s)$ which are located below and
the first Riemann sheet for all the channels located above. More specifically,
for $\operatorname{Re}[s]$ $\subset(s_{th,n},s_{th,n+1})$ we consider the
$(n+1)$-Riemann sheet for the whole function $\Pi(s)$ as defined as%
\begin{equation}
\label{rs4}
\Pi_{(n+1)}(s)=\sum_{j=1}^{n}\Pi_{j,II}(s)+\sum_{j=n+1}^{N}\Pi_{j}(s)\text{ .}%
\end{equation}
The prescription does not mean that there are not interesting poles on other
sheets (see e.g.~Ref.~\cite{prd93p014002}), but that those characterizing the resonance(s) is
(are) typically in one of the $N$ sheets above. As a last remark, while in
the first Riemann sheet $\Pi_{1}(s)=\Pi(s)$ does not have any pole, this is
not the case for other Riemann sheets. When searching for the poles of
$s-m_{0}^{2}+\Pi_{(n+1)}(s)=0$, besides the poles describing the property(s)
of resonance(s), other poles due to the form factor can emerge. In our work, we
could not find any of these spurious poles in any of the studied Riemann
sheets: it means that those poles are safely far from the real axis to have
any physical significance. For completeness, we further study this problem and
test a different form factor in Appendix \ref{AA}, to which we refer to for more
details.


\section{\label{description}The $\psi(4160)$}

\begin{table}
\centering
\begin{tabular}{c|c|c}
\hline
\vspace*{-3mm}&\\
 $j$&$(m_1m_2)_j$ & Th (MeV)\\
\hline
\vspace*{-3mm}&\\
1&$D^0\bar{D}^0$	&3729.66\\[1pt]
2&$D^+D^-$	&3739.18\\[1pt]
3&$D^0\bar{D}^{*0}+c.c.$	&3871.68\\[1pt]
4&$D^+D^{*-}+c.c.$&3879.85\\[1pt]
5&$D_s^+D_s^-$	&3936.54\\[1pt]
6&$D^{*0}\bar{D}^{*0}$	&4013.70\\[1pt]
7&$D^{*+}D^{*-}$	&4020.52\\[1pt]
8&$D_s^+D_s^{*-}+c.c.$	&4080.4\\[1pt]
9&$D_s^{*+}D_s^{*-}$	&4224.2\\[1pt]
\hline
\end{tabular}
\caption{\label{channels} Meson-meson internal loops and respective thresholds.}
\end{table}

We consider the vector charmonium $\psi(4160)$. The system includes a $\bar{c}c$ seed state with quantum numbers $2\ ^3D_1$ (the next radial excitation of the $\psi(3770)$), dressed by the meson-meson loops in Table \ref{channels}, which include pseudoscalar (P) and vector (V) fields. Clearly, all loops are on shell except the $\dsds$, whose threshold falls into the width of the $\psi(4160)$ ($4216\lesssim [m_\psi+\Gamma_\psi/2]\lesssim 4236$ MeV). With the definitions $\psi:=\psi(4160)$, $P:=D^0,D^+,D_s^+$, and $V:=D^{*0},D^{*+},D_s^{*+}$, three types of 3-vertex are involved, viz. $\psi PP$, $\psi PV$, and $\psi VV$. The corresponding Lagrangian densities are taken as: 
\be
\label{lagpp}
\L_{\psi PP}=ig_{\psi PP}\ \psi_\mu\Big(\d^\mu P\bar{P} - \d^\mu \bar{P}P\Big)+h.c.\ ,
\ee
\be
\label{lagpv}
\L_{\psi PV}=g_{\psi PV}\ \tilde{\Psi}_{\mu\nu}P\bar{V}^{\mu\nu}+h.c.\ ,\ \mathrm{and}
\ee
\be
\label{lagvv}
\L_{\psi VV}=ig_{\psi VV}\ \Psi_{\mu\nu}\Big(V^{\mu}\bar{V}^{\nu} - V^{\nu}\bar{V}^{\mu} \Big) + h.c.\ ,
\ee
with the definitions
\be
\label{redpsi}
\Psi^{\mu\nu}=\partial^\mu\psi^\nu-\partial^\nu\psi^\mu,\ \ \tilde{\Psi}_{\mu\nu}=\frac{1}{2}\epsilon_{\mu\nu\alpha\beta}\Psi^{\alpha\beta}.\\
\ee
From Eqs.~\eqref{lagpp}-\eqref{redpsi}, we obtain the following amplitudes, in the $\psi$ rest frame, i.e.~$s=m_\psi^2$,
\be
\label{app}
\mathcal{V}_{\psi\to(PP)_j}=\frac{4}{3}g^2_{\psi PP}\ \vec{k}_j^2\ ,
\ee
\be
\label{apv}
\mathcal{V}_{\psi\to(PV)_j}(s)=\frac{1}{3}g^2_{\psi PV}\ s\Big(3m_V^2+2\vec{k}_j^2\Big)\ ,
\ee
\be
\label{avv}
\mathcal{V}_{\psi\to(VV)_j}(s)=\frac{16}{3}g^2_{\psi VV}\ s\bigg( \frac{\vec{k}_j^4}{m_V^4}+2\frac{\vec{k}_j^2}{m_V^2}\bigg)\ .
\ee

\begin{figure}
\centering
\begin{tabular}{c}
\hspace*{-0.6cm}\resizebox{!}{230pt}{\includegraphics{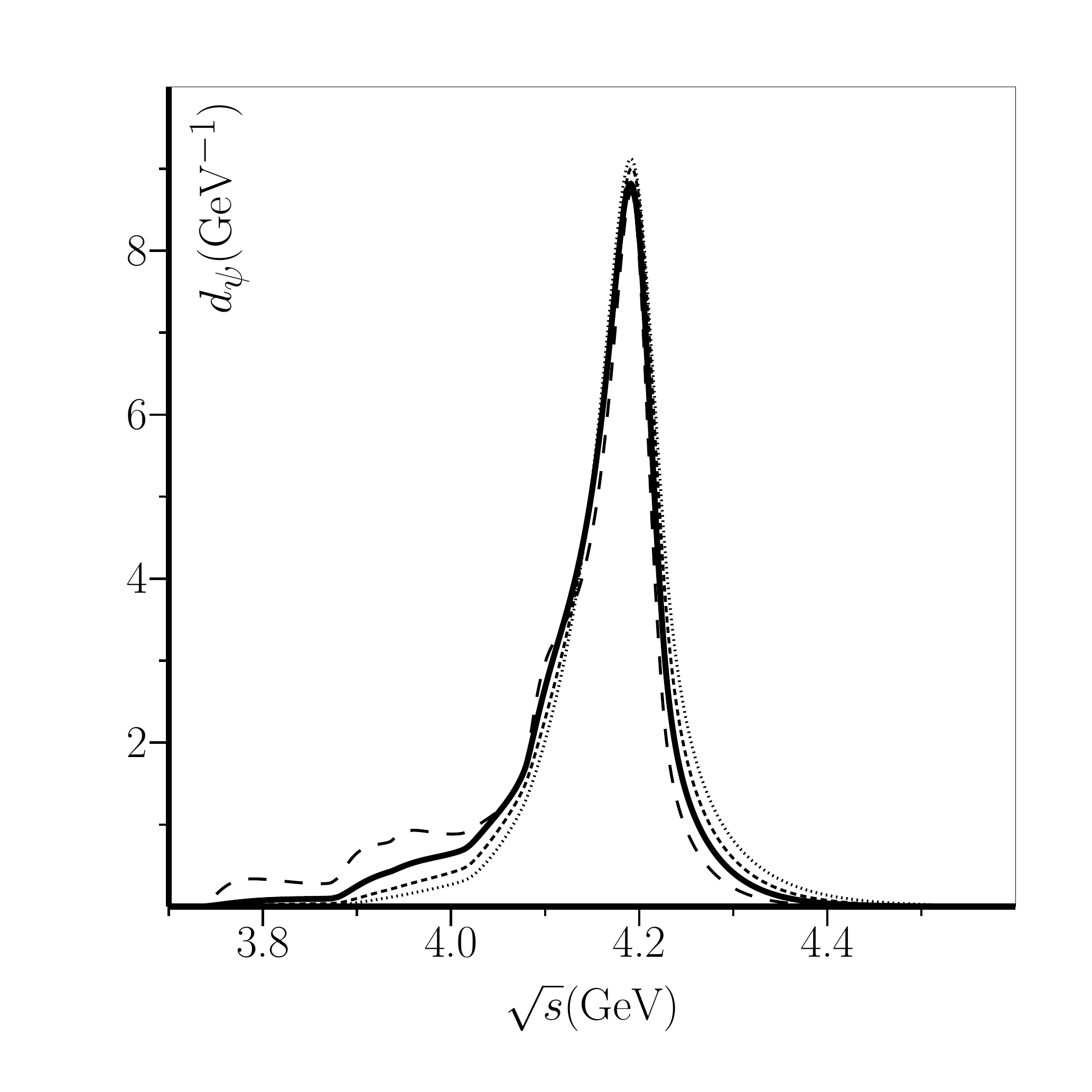}}
\end{tabular}
\caption{\label{psi} Spectral function of the $\psi(4160)$, as a function of the ``running'' mass $m_\psi=\sqrt{s}$, varying with $\Lambda$: long dashed line, $\Lambda$=400 MeV; solid line, $\Lambda$=450 MeV; short-dashed line, $\Lambda$=500 MeV; dotted line, $\Lambda$=550 MeV. In each case, the peak is at about $4.191$ GeV.}
\end{figure}

Having the amplitudes in Eqs.~\eqref{app}-\eqref{avv} inserted in Eq.~\eqref{amp}, the spectral function for the $\psi(4160)$ in Eq.~\eqref{sfi} or \eqref{sf} is fully defined, except for five free parameters: the seed mass $m_0$ in Eq.~\eqref{prop}, the cutoff parameter $\Lambda$ in Eq.~\eqref{ff}, and the partial coupling constants $g_{\psi PP}$, $g_{\psi PV}$, and $g_{\psi VV}$ entering in the amplitudes. Four of these parameters are constrained by four experimental quantities in Ref.~\cite{pdg}: first, we impose $m_0$ to be such that the mass of the peak in the spectral function \eqref{sfi} or \eqref{sf} is equal to the average mass of the $\psi(4160)$, i.e.~$m_{peak}\simeq 4191$ MeV; secondly, for a fixed $\Lambda$, we constrain the value of the three partial couplings by imposing the total width of the peak $\Gamma_{peak}=\sum_j^N\Gamma_j(m^2_{peak})$ to fall in the average width for the $\psi(4160)$, i.e. $\Gamma_{peak}\simeq 70$ MeV, and the ratios $\Gamma_{D\bar{D}}(m^2_{peak})/\Gamma_{D^*\bar{D}^*}(m^2_{peak})\simeq 0.02$ and $\Gamma_{D^*\bar{D}+c.c.}(m^2_{peak})/\Gamma_{D^*\bar{D}^*}(m^2_{peak})\simeq 0.34$, using Eqs.~\eqref{gamma}-\eqref{ff} and \eqref{app}-\eqref{avv}, assuming flavor independent decays. With this setup, we are left with only one free parameter, the cutoff $\Lambda$. In Fig.~\ref{psi} we show the variation of the line-shape of the $\psi(4160)$, given by Eq.~\eqref{sfi} or \eqref{sf}, for $\Lambda$=400, 450, 500, and 550 MeV, having the remaining parameters listed in Table \ref{pvar}. From Fig.~\ref{psi} it can be seen that, in the energy region of our problem, i.e.~around 4.2 GeV, the qualitative line-shape is weakly dependent on the specific value of the cutoff. Since we do not include the $\psi(4040)$ as a second seed, the spectral function at lower energies is inaccurate. In Table \ref{pvar} we also show the pole position corresponding to the $\psi(4160)$, for each $\Lambda$ value, computed through Eqs.~\eqref{poles}-\eqref{rs4}. For each case, only one pole is found, coming from the seed state.  The seed mass is generally lower than the physical mass of the $\psi$, showing that the ``unquenching'' pulls the seed pole upwards.

\begin{table}
\centering
\begin{tabular}{c|c|c|c|c|c}
\hline
\vspace*{-3mm}&&&&\\
 $\Lambda$ (MeV) & $m_0$ (MeV)& $g_{\psi PP}$ &$g_{\psi PV}$ (GeV$^{-1}$)&$g_{\psi VV}$&Pole (MeV)\\
\hline
\vspace*{-3mm}&&&&\\
400	&4127  &52.9&6.30&4.07&$4198.3-i27.2$\\[1pt]
450	&4153.6&23.8&4.04&3.76&$4199.2-i32.7$\\[1pt]
500	&4170  &12.5 &2.72&3.29&$4200.0-i36.4$\\[1pt]
550     &4180  &7.38&1.94&2.84&$4198.1-i40.2$\\[1pt]
\hline
\end{tabular}
\caption{\label{pvar} Variation of the free parameters with $\Lambda$, and pole positions for the $\psi(4160)$ (see text for details).}
\end{table}

\begin{figure}
\centering
\begin{tabular}{c}
\hspace*{-0.6cm}\resizebox{!}{230pt}{\includegraphics{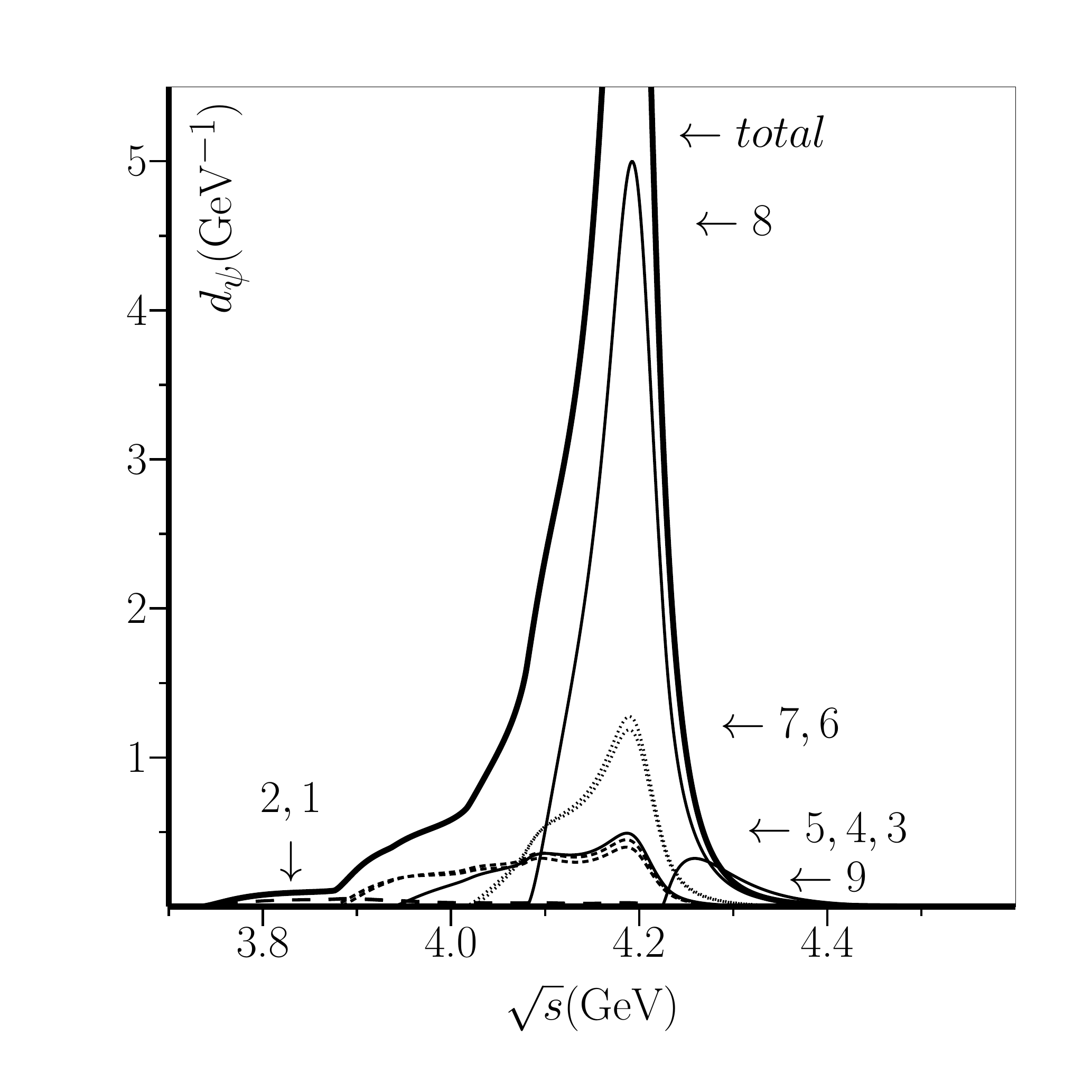}}
\end{tabular}
\caption{\label{partialsf} Total and partial spectral functions for $\Lambda=450$ MeV. Numbers correspond to channels in Table \ref{channels}. Solid bold line: total; solid lines: channels 5, 8, and 9, respectively; dotted lines: channels 6 (down) and 7 (up); short-dashed lines: channels 3 (down) and 4 (up); long-dashed lines: channels 1 (down) and 2 (up).}
\end{figure}

In Fig.~\ref{partialsf} we show the partial spectral functions in Eq.~\eqref{sfp}, for $\Lambda=450$ MeV. The peak position is approximately the same for each partial spectral function, although the specific form of the line-shape varies, as function of the kinematics and amplitude. In fact, this can be expected from Eq.~\eqref{sf}, since the denominator is common for all channels, and the numerator is a regular function. Therefore, the one-loop effect alone cannot reproduce any mass shifting in a particular channel only, as it was already concluded in Ref.~\cite{1810.03532}. 

The formalism allows for the inclusion of off-shell loops, that also influence the $\psi(4160)$, namely the $D\bar{D}_1+c.c.$ and $D\bar{D}'_1+c.c.$ loops, however we exclude them for simplicity, to avoid the introduction of additional free parameters through the partial couplings. Their inclusion is not expected to be, anyhow, significant.


\section{\label{interference}The $Y(4260)$}
 
In the previous section, we described the resonance $\psi(4160)$ using an unitary model with internal loops, chosen according to the OZI-allowed rule. In this section, we study the cross section of the $\psi(4160)$ in the OZI-suppressed channel $J/\psi f_0(980)$, that subsequently decays into $J/\psi \pi^+\pi^-$. In Fig.~\ref{psi}, we have seen that the specific choice of the cutoff parameter does not change the general result. Here, we take the value $\Lambda=450$ MeV. 


\subsection{\label{direct}Direct decay: $\psi\to J/\psi f_0(980)$ }
\begin{figure}
\centering
\begin{tabular}{c}
\hspace*{-0.6cm}\resizebox{!}{230pt}{\includegraphics{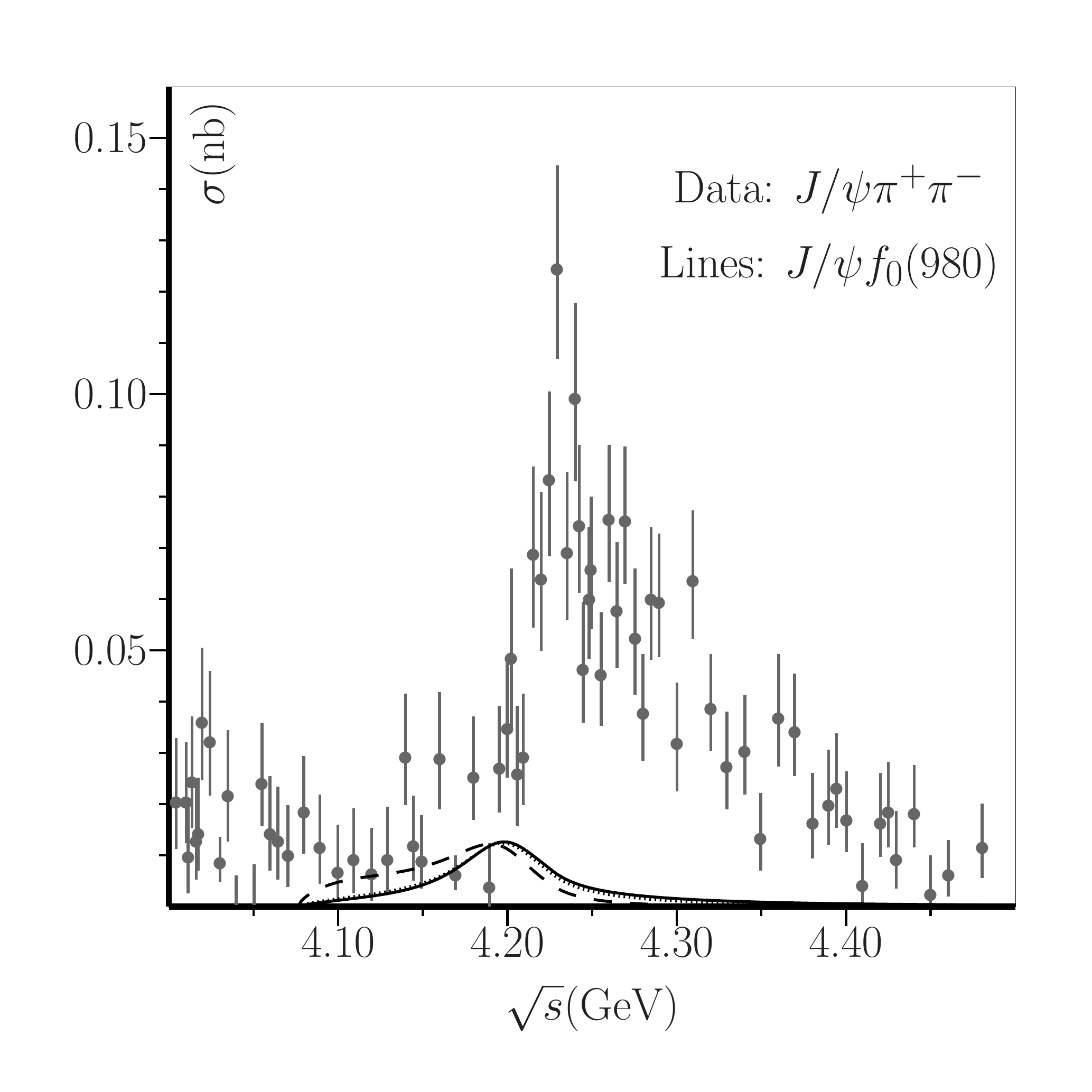}}
\end{tabular}
\caption{\label{jppnormal} Cross section for the process $e^+e^-\to \psi\to J/\psi f_0(980)\to J/\psi \pi^+\pi^-$. Data points from BESIII in Ref.~\cite{prl118p092001}. Dashed line: $\tilde{\Lambda}=450$ MeV; dotted line: $\tilde{\Lambda}=1$ GeV; solid line: $\tilde{\Lambda}=10$ GeV. No description of data, using Eq.~\eqref{lagjf}, is possible.}
\end{figure}

Let us include the new decay channel $J/\psi f_0(980)$ exactly in the same way as the decay channels in Table \ref{channels}, by defining the new interaction Lagrangian as
\be
\label{lagjf}
\mathcal{L}_{\psi J f_0}=g_{\psi J f_0}\ \Psi_{\mu\nu} J^{\mu\nu} f_0 +h.c.\ ,
\ee  
with $J\equiv J/\psi$ and $f_0\equiv f_0(980)$. With $s=m_\psi^2$, it leads to the amplitude
\be
\label{ajf}
\mathcal{V}_{\psi\to J f_0}(s)=\frac{4}{3}g_{\psi J f_0}^2\ s\bigg[2k^2(s,m_{J},m_{f_0})+3m_{J/\psi}^2\bigg].
\ee
Furthermore, we consider that, due to the different decay mechanism, and participation of light mesons, the cutoff parameter relative to channel $J/\psi f_0(980)$, that we define as $\tilde{\Lambda}$, may differ from the general one $\Lambda$. Now, we compute the cross section for the $e^+e^-\to\psi\to J/\psi f_0(980)$ production, through
\be
\label{cs}
\sigma_{e^+e^-\to J f_0}(\sqrt{s})=\frac{\pi}{2\sqrt{s}}\ g_{\psi e^+e^-}^2\ d_{\psi\to J f_0}(\sqrt{s})\ ,
\ee
using Eq.~\eqref{sfp}, with 
\be
\label{gdirect}
\Gamma_{\psi\to J/\psi f_0}(s)=\frac{k(s,m_{J},m_{f_0})}{8\pi s}\mathcal{V}_{\psi\to J f_0}(s)f_{\tilde{\Lambda}}^2(k^2(s,m_{J},m_{f_0}))\ .
\ee
 The coupling $g_{\psi e^+e^-}$, in Eq.~\eqref{cs}, may be estimated from the experimental decay $\Gamma_{\psi(4160)\to e^+e^-}\simeq 0.44\pm 0.22$ keV \cite{pdg}, using
\be
\label{csll}
\Gamma_{\psi(4160)\to e^+e^-}(s)=g_{\psi e^+e^-}^2\frac{4}{3}\frac{k(s,m_{e})}{8\pi s}\Big(s+2m_e^2\Big)\ .
\ee
It gives $g_{\psi e^+e^-}\simeq 1.989\times 10^{-3}$. In Fig.~\ref{jppnormal} we compare the theoretical cross section in Eq.~\eqref{cs} with the $J/\psi\pi^+\pi^-$ data in Ref.~\cite{prl118p092001}, by adjusting the parameters $\tilde{\Lambda}$ and $g_{\psi J f_0}$ as following:
\be
\label{pan}
\begin{split}
&\tilde{\Lambda}=450\ \mathrm{MeV\ ,\ } g_{\psi J f_0}\simeq 0.110 \mathrm{\ GeV}^{-1}\ ,\\
&\tilde{\Lambda}=1\ \mathrm{GeV\ ,\ } g_{\psi J f_0}\simeq 0.054\mathrm{\ GeV}^{-1}\ ,\\
&\tilde{\Lambda}=10\ \mathrm{GeV\ ,\ } g_{\psi J f_0}\simeq 0.051\mathrm{\ GeV}^{-1}\ .\\
\end{split}
\ee
Since the value of the parameter $\tilde{\Lambda}$ is not
known, we test three different scenarios: a `small' $\tilde{\Lambda}=450$ MeV
(similar to the value of $\Lambda),$ an intermediate value $\tilde{\Lambda
}=1$ GeV (typical when light mesons are involved), and a very large value
$\tilde{\Lambda}=10$ GeV (in practice, `infinite'.) In each case, the coupling
constant $g_{\psi Jf_{0}}$ is a test value used to generate Fig.~\ref{jppnormal}: the
corresponding cross section has a peak at the mass of $\psi(4160)$, that
\textit{has not been seen} in experiments. Hence, the values quoted in\ Eq.~\eqref{pan} can be also seen as an estimate of the maximal value for such couplings
(since, if it were sizably larger, one would have seen it in experimental
data).

We observe that, independently of the parameters set, the peak in the cross section always comes at about 4.19 GeV, i.e., at the mass of the $\psi(4160)$, which is determined by the corresponding underlying pole (cf.~Table \ref{pvar}). Then, it is not possible that the interaction in Eq.~\eqref{lagjf} describes the data in the $J/\psi\pi^+\pi^-$ decay mode: the peak is too small and placed at too small $\sqrt{s}$. The coupling parameters in \eqref{pan} are  illustrative of how much the direct decay $\psi\to J/\psi f_0(980)$, that occurs through gluon emission and subsequent conversion into quark-antiquark pairs (OZI-suppressed process), is suppressed. From Fig.~\ref{jppnormal} we conclude that the peak at about 4.23 GeV in the data cannot be described within the simple one-loop mechanism we presented so far. In the next section we explore a different production process for the $J/\psi f_0(980)$ that changes this result. 

\subsection{\label{inter} Loop-driven decay: $\psi\to\dsds\to\vpsif0$}

\begin{figure}
\centering
\begin{tabular}{c}
\hspace*{-0.6cm}\resizebox{!}{75pt}{\includegraphics{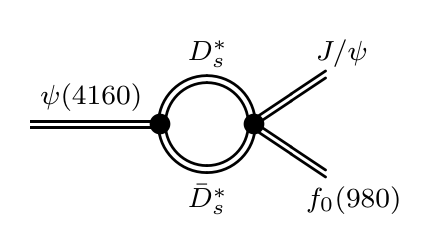}}
\end{tabular}
\caption{\label{4vertex} Final-state interaction, which accounts for the mass-shift effect in Fig.~\ref{main}.}
\end{figure}

Let us consider the production process $\psi\to\dsds\to\vpsif0$, according to the scheme in Fig.~\ref{4vertex}. Such interaction is possible because the quark content of the $\dsds$ and $\vpsif0$ is the same, i.e. $\{c,\bar{c},s,\bar{s}\}$, given that the $f_0(980)$ has a sizable $\bar{s}s$ component in its wave-function. Furthermore, the pole corresponding to the $\psi(4160)$ (cf.~Table \ref{pvar}) is very close, yet below the $\dsds$ threshold (see line $9$ in Fig.~\ref{partialsf}). The fact that the pole is below threshold makes the $\dsds$ mostly off shell, which means that, while a decay through a string-breaking mechanism, i.e. OZI-allowed, is strongly favored, the fact that the phase space for the decay is very limited enhances the possibility of an internal quark recombination into a lighter meson-meson system.     

\begin{figure}
\centering
\resizebox{!}{230pt}{\includegraphics{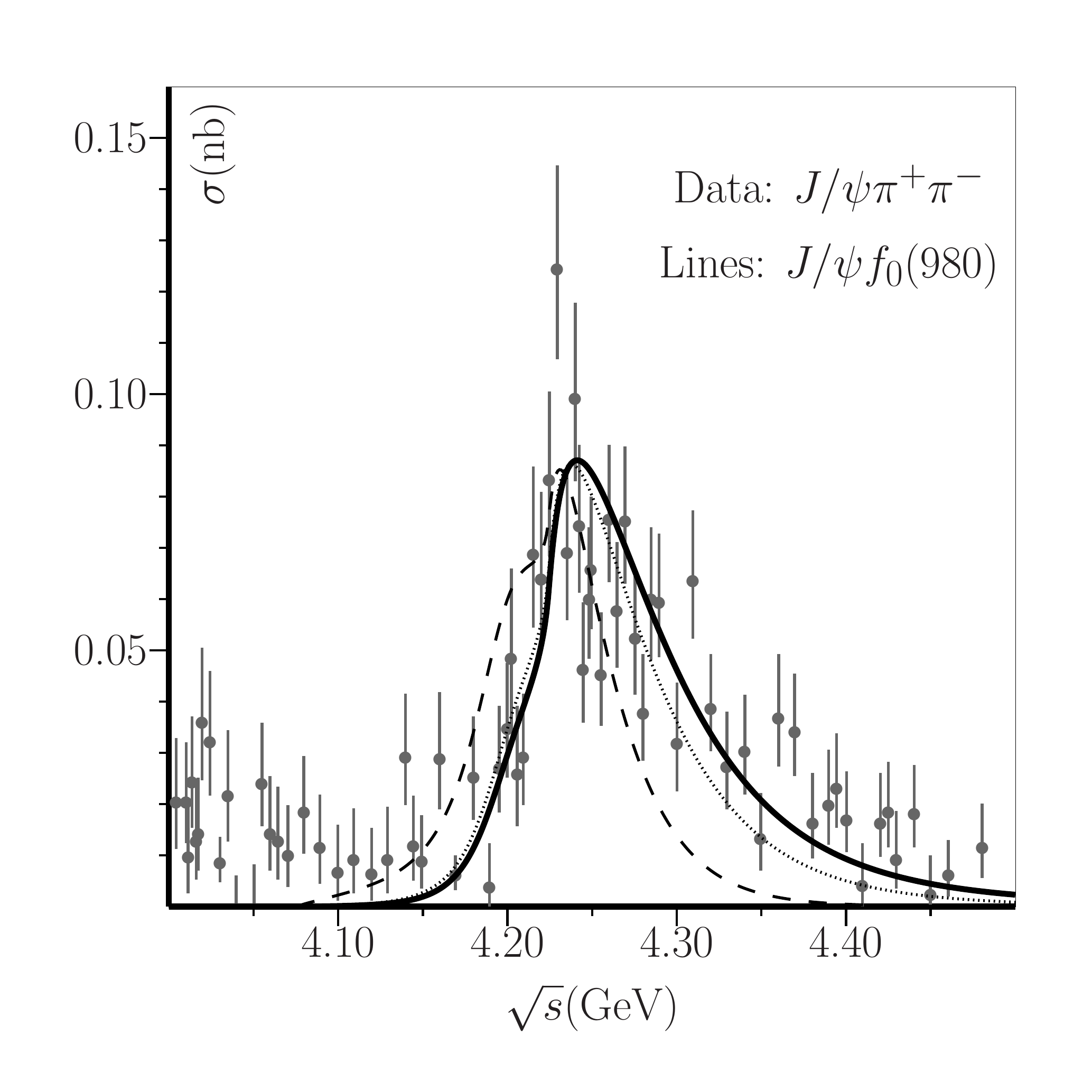}}
\caption{\label{main} Cross-section for $e^+e^-\to\psi(4160)\to\dsds\to J/\psi f_0(980)$, using the loop-driven decay in Fig.~\ref{4vertex}, compared with the experimental cross-section $e^+e^-\to J/\psi \pi^+\pi^-$ in Ref.~\cite{prl118p092001}. Dashed, dotted and solid line: $\tilde{\Lambda}=450$ MeV, 1 GeV, and 10 GeV, respectively. See text for details.}
\end{figure}

\begin{figure}
\centering
\resizebox{!}{230pt}{\includegraphics{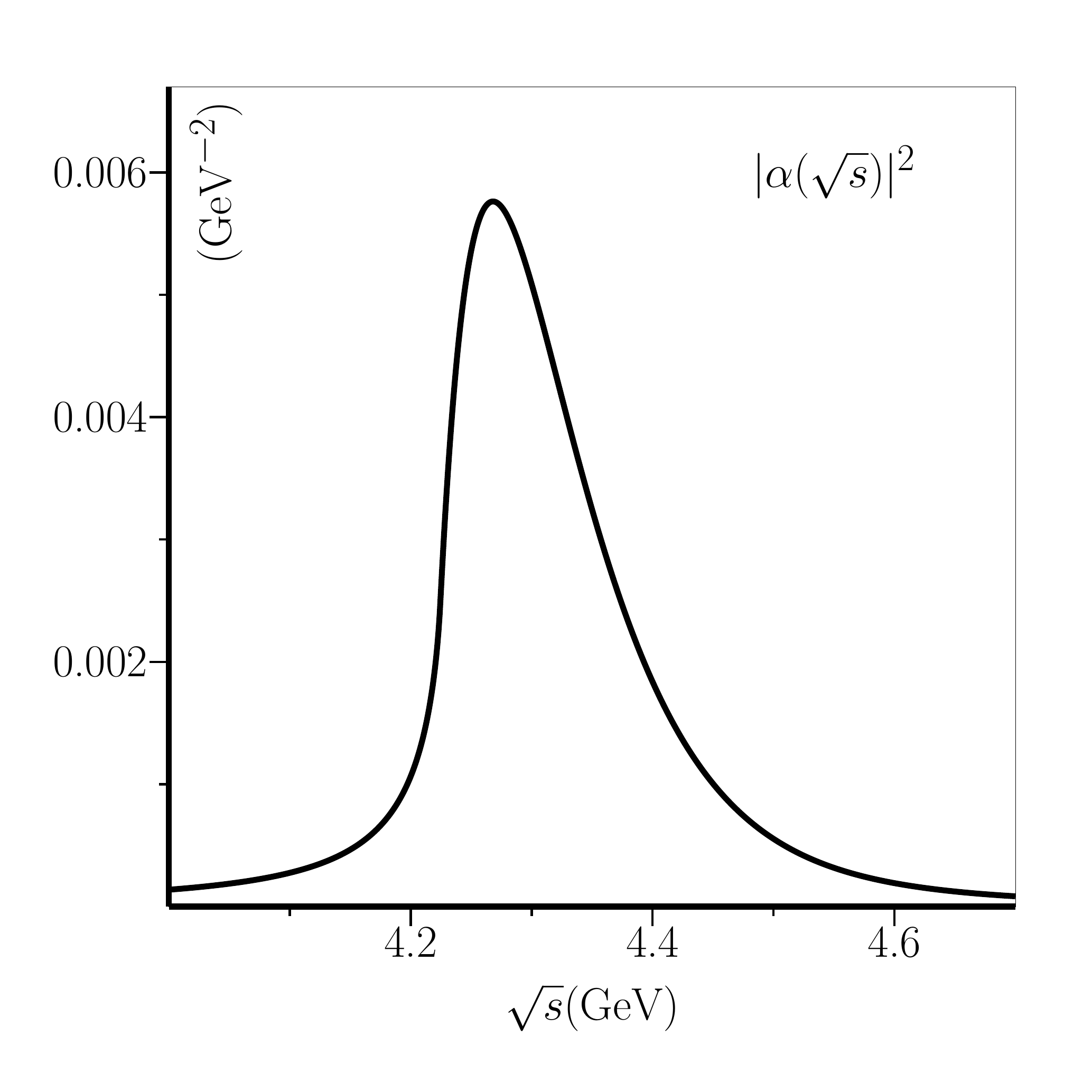}}
\caption{\label{alsq} Modulus square of the loop in Eq.~\eqref{alpha}, that acts as an energy dependent coupling. We can observe that the function arises around 4.2 GeV, which is the reason why the amplitude in Eq.~\eqref{amp4} and subsequent equations are enhanced at that energy.}
\end{figure}

The 3-vertex $\psi\dsds$ interaction in Fig.~\ref{4vertex} is given by Eq.~\eqref{lagvv}, providing that $V\bar{V}=\dsds$, and the 4-vertex $\dsds\vpsif0$ interaction is given by the Lagrangian 
\footnote{\label{foot}The transition $\dsds\to J f_{0}$ is modelled by Eq.~\eqref{lagvv4}: this is the Lagrangian whose interaction
term has the least number of derivatives and represents a suitable way to
parametrize this transition with only one free parameter, the coupling
$\lambda$. The shape of the corresponding cross-section for the $J/\psi\pi
^{+}\pi^{-}$ production (see later on and in Fig.~\ref{main}), does not depend on the
value of the constant $\lambda$ (only the height does). As we shall see, a
peak at about $4.23$ GeV, just where $Y(4260)$ sits, emerges (independent on
$\lambda$), thus the possibility to describe this state as a shifted peak of
$\psi(4160)$ seems appealing. The possible numerical value(s) of $\lambda$ is
(are) obtained by requiring that the height of the cross-section is in
agreement with the data, see below for details. Moreover, the contribution of
the small direct decay studied in the previous section generates an
interference phenomenon which improves the description of data. In the future,
the inclusion of more terms that describe the $D_{s}^{\ast}D_{s}^{\ast
}\rightarrow J f_{0}$ transition (as well as other subleading but
possible mechanisms leading to $J/\psi\pi^{+}\pi^{-}$ in the final state)
would be interesting, but the proliferation of coupling constants as well as
the technical involvement would make such a task valuable once much more
precise data will be available.} 
\be
\label{lagvv4}
\L_{\dsds\to Jf_0}=i\lambda \Big(D_{s\mu}^*\bar{D}_{s\nu}^* - D_{s\nu}^*\bar{D}_{s\mu}^* \Big)f_0J^{\mu\nu} + h.c.\ ,
\ee
with the definitions in Eq.~\eqref{redpsi}. A detailed calculation of the diagram in Fig.~\ref{4vertex} is given by the product of the 3-vertex amplitude \eqref{lagvv}, the $\dsds$ loop integral, and the 4-vertex amplitude \eqref{lagvv4}. The result is an effective amplitude similar to Eq.~\eqref{ajf}, but where in the place of the coupling strength $g_{\psi J f_0}$, it comes a new effective energy dependent coupling that includes the $\dsds$ loop, which is given by
\be
\label{alpha}
\alpha(\sqrt{s})=\frac{\lambda \Pi_{D_s^*D_s^*}(s)}{s}\ ,
\ee
where $\Pi_{\dsds}(s)$ is the loop function, that includes the coupling strength $g_{\psi\dsds}$ of the left vertex in Fig.~\ref{4vertex}, $\lambda$ is the coupling in \eqref{lagvv4}, and $s$ regularizes the dimensions. The total $\psi\to J/\psi f_0(980)$ amplitude is written as 
\be
\label{amp4}
\mathcal{\tilde{V}}_{\psi\to J f_0}(s)=\frac{4}{3}|\alpha(\sqrt{s})\pm g_{Jf_0}|^2 s\bigg[2k^2(s,m_{J},m_{f_0})+3m_{J}^2\bigg]\ .
\ee
The complex coupling term $|\alpha(\sqrt{s})\pm g_{Jf_0}|^2$ includes the pure loop-driven process in Fig.~\ref{4vertex}, the direct process, and an additional interference term between the two. The sign $\pm$ represents the case in which Re($\alpha(\sqrt{s})$) and $g_{Jf_0}$ have the same sign ($+$) or opposite sign ($-$). We shall see below that only the opposite sign leads to a good comparison with the data. The amplitude \eqref{amp4} enters into the new decay width as
\be
\label{gamtilde}
\tilde{\Gamma}_{\psi\to J f_0}(s)=\frac{k(s,m_{J},m_{f_0})}{8\pi s}\times\mathcal{\tilde{V}}_{\psi\to J f_0}(s)\times e^{-\vec{k}^2(s,m_{J},m_{f_0})/\tilde{\Lambda}^2}\ ,
\ee 
where we use the notation $\tilde{\mathcal{V}}$ and $\tilde{\Gamma}$ to distinguish from the direct process in Eqs.~\eqref{ajf} and \eqref{gdirect}. As in the direct decay case, we note that the cutoff $\tilde{\Lambda}$, in the above equation, is different than the one used for the OZI-allowed loop vertices. The computation of the cross section for the $e^+e^-\to \psi\to J/\psi f_0(980)$ production is done through Eqs.~\eqref{sfp}, \eqref{cs}, \eqref{csll}, and \eqref{alpha}-\eqref{gamtilde}. The total cross section in channel $\vpsif0$ will be
\be
\label{csf}
\tilde{\sigma}_{\psi\to J f_0}=\sigma_{\psi\to J f_0}^\mathrm{direct}+\sigma_{\psi\to\dsds\to J f_0}^\mathrm{loop-driven}+\sigma^\mathrm{interference}\ .
\ee
In Fig.~\ref{main} we plot the total cross section $\tilde{\sigma}_{\psi\to\vpsif0}$ in the above equation for the following parameters:
\be
\label{pan2}
\begin{split}
&\tilde{\Lambda}=450\ \mathrm{MeV\ ,\ } \lambda=15.2 \mathrm{\ GeV}^{-1}\ ,\\
&\tilde{\Lambda}=1\ \mathrm{GeV\ ,\ } \lambda=6.1\mathrm{\ GeV}^{-1}\ ,\\
&\tilde{\Lambda}=10\ \mathrm{GeV\ ,\ } \lambda=4.9\mathrm{\ GeV}^{-1}\ ,\\
\end{split}
\ee
where each $\lambda$ is adjusted for comparison with data in Ref.~\cite{prl118p092001}. We use the corresponding $g_{J f_0}$ values as in Eq.~\eqref{pan}. The results are depicted for the case where the interference between the direct and the loop-driven processes is negative, i.e., minus sign in Eq.~\eqref{amp4}, which are those that describe data the best. The results are in very good agreement with data if we allow a larger value for $\tilde{\Lambda}$. The most striking feature of the Fig.~\ref{main} is that the peak clearly shifts from its position, around 4.19 GeV, to about 4.23 GeV, matching the structure of the $Y(4260)$. The function $\alpha(\sqrt{s})$ is responsible for this shift. In fact, from Fig.~\ref{partialsf}, we can already see that channel $\dsds$ (line 9) reaches its maximal value around 4.26 GeV. In Fig.~\ref{alsq}, we draw the function $|\alpha(\sqrt{s})|^2$, for $\tilde{\Lambda}=10$ GeV, and $\lambda=4.9$ GeV$^{-1}$ (cf.~\eqref{pan2}). It reaches a maximal value for $\sqrt{s}\simeq 4.27$ GeV at about $9.10\times 10^{-3}$ GeV$^{-2}$, which is much smaller than the square of the couplings in Table \ref{channels}. The maximal width in Eq.~\eqref{gamtilde} comes at
\be
\tilde{\Gamma}_{\psi\to J/\psi f_0(980)}^\mathrm{\ max}(s\simeq 4.30^2\ \mathrm{MeV}^2)\simeq 7.21\ \mathrm{MeV}\ , 
\ee 
which we determine graphically. However, at the physical mass of the $\psi(4160)$, it is
\be
\tilde{\Gamma}_{\psi\to J/\psi f_0(980)}(s\simeq 4.191^2\ \mathrm{MeV}^2)\simeq 0.29\ \mathrm{MeV}\ ,
\ee
a value that is close to the upper limit given in Ref.~\cite{pdg} of $3\times 10^{-3}\times 70$ MeV $=0.21$ MeV. For simplicity, we do not include Eq.~\eqref{gamtilde} in the denominator of Eq.~\eqref{sf}. As a consequence, there is a small violation of unitarity of about 1.5$\%$, which we consider to be negligible, thus confirming {\it a posteriori} our approximation. In fact, for consistency, if the $\vpsif0\dsds$ 4-vertex interaction would be included in the denominator, all other 4-vertex interactions should also be included. This would unnecessarily increase the complexity of our problem, without changing the outcomes sizably. 

The theoretical peak in Fig.~\ref{main}, which is the main result of our study, is actually a variation of the $\psi(4160)$ itself, when {\it shifted}  by the influence of the loop-driven effect. The direct decay contribution, with peak at the nominal mass of the $\psi(4190)$, is still present, but its partial cross section to channel $J/\psi\pi^+\pi^-$ is, on the one hand, very small (see Fig.~\ref{jppnormal}), and on the other hand, it suffers negative interference with the loop-driven decay, in such a way that it is dominated by it. Since the $f_0(980)$ also has a component of $u$ and $d$ quarks (hence its strong decay to $\pi\pi$), a contribution of other 4-vertices, e.g. the $\vpsif0 D^*\bar{D}^*$, is also present, but in those cases, the corresponding $\alpha$ function \eqref{alpha} has its peak around the threshold mass of the corresponding OZI-allowed meson-meson pair, becoming very small around 4.23 GeV. On the other hand, the peak at about 4.23 GeV in Fig.~\ref{main} is also present in the other OZI-allowed channels, that couple to the $\dsds$ channel through the same sort of final state interaction as in Fig.~\ref{4vertex}, but it is not seen in those channels due to the dominance of the direct process in such cases. We remark that, the existence of the structure at 4.23 GeV is, within our approach, intrinsically related to the existence of an off-shell threshold very close to the pole of the $\psi(4160)$.

In this work, we consider the resonance $f_{0}(980)$ as an
intermediate state for the $D_{s}^{\ast}D_{s}^{\ast}\rightarrow J/\Psi\pi
^{+}\pi^{-}$ production for mainly three reasons: (i) it couples strongly to
kaons, assuring a strong coupling to a $\bar{s}s$ pair, necessary in the
formation process (indeed, the $f_{0}(980)$ is often interpreted as a
four-quark object in which $\bar{s}s$ enters in its wave function); (ii) it
couples strongly to pions, necessary for the production of a $\pi\pi$ pair in
the final state; (iii) it is kinematically favoured, since $m_{J/\Psi
}+m_{f_{0}(980)}<m_{\psi(4160)}$.

Yet, there are other resonances of the $f_{0}$ type that can also contribute
to the decay channel and, in principle, one should perform the sum over all of
them: the light state $f_{0}(500)$ \cite{pr658p1} couples strongly to pions
and is kinematically even more favored than $f_{0}(980)$, but its coupling to
$\bar{s}s$ is not known and could be not large if $f_{0}(500)$ is
predominantly nonstrange; the state $f_{0}(1370),$ which couples to both
$\bar{s}s$ and pions; finally the coupling to $f_{0}(1500)$ and also
$f_{0}(1710)$ could have a non-negligible influence. Note, $f_{0}(1360),$
$f_{0}(1500),$ and $f_{0}(1710)$ are kinematically not allowed for an on-shell
decay, but they clearly contribute as virtual state to the final $J/\Psi
\pi^{+}\pi^{-}$product. 

The PDG does not present yet a fit or average for the contribution of
$f_{0}(980)$ to the final state $J/\Psi\pi^{+}\pi^{-}$ (it is surely seen and
sizable, yet the fraction is unknown). The experiment in\ Ref.~\cite{prd86p051102} finds
that this ratio is $0.17\pm0.13.$ Our argumentation suggests that it should be
larger. Future experimental results on this ratio would be very welcome.

While the detailed inclusions of all these $f_{0}$ resonances is left for
future works (one would need a way to estimate the coupling to all these
states and also take into account possible interference phenomena), it should
be noted that the main idea presented here, the $D_{s}^{\ast}D_{s}^{\ast}$
loop as intermediate state, would be very similar in those channels as well
and the $J/\Psi\pi^{+}\pi^{-}$ would peak at very similar values of $\sqrt
{s}\simeq4.23$ GeV Hence, the study of $f_{0}(980)$ presented in this work
represents the prototype for all other $f_{0}$ resonances. 

In the Appendix \ref{AA} we discuss the possibility of using a different cutoff-function, namely with a quadripolar form. While a shift in the peak is still seen, the result is less striking, and thus we conclude that the exponential function works better for the current problem.

Other discussions, concerning other contributions to the $J/\psi\pi^+\pi^-$ final state, a comment on the experimental result in channel $DD^*\pi$ in Ref.~\cite{prl122p102002}, and on the cross section for the direct decay $\psi\to\dsds$, may be found in Appendices \ref{AB}, \ref{AC}, and \ref{AD}, respectively.

\section{\label{conclusion}Conclusion and Perspectives}

We have presented a novel possible interpretation of the $Y(4260)$, which emerges from a loop-driven decay involving the $\dsds$ and the $\vpsif0$ meson pairs, with only one underlying pole, that corresponds to the $\psi(4160)$ resonance. The effect is manifest due to the close proximity of the pole to the mostly closed threshold $\dsds$. While the coupling between the $\psi(4160)$ and this OZI-allowed channel is high, the lack of phase space for the decay enhances the possibility of recombination of the quark content of the $\dsds$ into an OZI-suppressed decay mode, viz. $\vpsif0$,  with a lot of phase space available. Furthermore, a negative interference between the loop-driven decay and the direct decay, enhances the peak arising at about 4.23 GeV.  The conditions for the formation of the $Y(4260)$ structure are, therefore, very precise.  Without changing the position of the $\psi(4160)$ pole, the effective line-shape $d_{\psi\to J/\psi f_0(980)}$, and consequently the cross section, undergoes an ``energy shift'' upwards, a result that we consider as quite remarkable, and that opens new possibilities to address the enigmatic $Y$ enhancements.

We note that we are not including here other $\psi$ resonances, such as the $\psi(4040)$ and the $\psi(4415)$, that surely have influence in a more comprehensive study. Notwithstanding the conclusions of other works, namely \cite{prd83p054021,prd93p014011}, the interference among different $\psi$ is, within the present approach, not necessary to explain the bulk of the structure seen in the data, viz.~the $Y(4260)$. The direct comparison made in Fig.~\ref{main} between the $J/\psi\pi^+\pi^-$ and the $J/\psi f_0(980)$ is not quantitatively strict. On the one hand, the $J/\psi\pi^+\pi^-$ may result from other decays, such as from the $Z_c(3900)^\pm\pi^\mp$. In fact, according to experiment \cite{pdg}, and also certain analysis \cite{0701002}, such contribution is significant. Its ratio w.r.t.~the total
$J/\Psi\pi^{+}\pi^{-}$ channel is $0.215\pm0.033\pm0.075$ (suppressed but not
negligible). On the other hand, the $f_0(980)$ also decays into $\pi^0\pi^0$ and to $KK$. The channels $J/\psi\pi^0\pi^0$ and $J/\psi K^+K^-$ \cite{prd97p071101} are, therefore, candidates for future studies of the $Y(4260)$. Furthermore, the $J/\Psi\pi^{+}\pi^{-}$ may result from
other scalar resonances, such as $f_{0}(500),$ $f_{0}(1370),$ $f_{0}(1500),$
and $f_{0}(1710)$. In future studies, one should repeat the calculation
performed in this work for all these channels and take properly into account
eventual interference effects. To this end, a model for the coupling to all
these scalar states is needed. Yet, the peak of this reaction is determined by
the $\dsds$ loop and would be very similar also when
including all these scalar states. Within the present effective Lagrangian approach, the orbital angular momentum is not explicit, however we consider the $\psi(4160)$ to be a dominantly $d$-wave state, in which case the $Y(4260)$ enhancement should also be in $d$-wave. A similar mechanism, involving the $s$-wave counterpart of the $\psi(4160)$, i.e.~the $\psi(4040)$, has been studied by one of us in Ref.~\cite{epjc79p98}, to explore the possible $Y(4008)$ enhancement. For the present work, other possible effects are the interference between $D\bar{D}_1+c.c.$ and $D\bar{D}_1'+c.c.$ loops and the $Z_c\pi$ channel. Such effects shall be, however, significantly smaller than the one in Fig.~\ref{4vertex}, since the corresponding thresholds, about 4.29 GeV, are far enough from the peak of the $\psi(4160)$. One should also notice that the actual mass of the $Y(4260)$ is now around 4.23 GeV, thus further away from the $DD_1$ threshold than what was initially measured. 
Another interesting mechanism, also involving $D\bar{D}_1+c.c.$ and $D\bar{D}_1'+c.c.$ loops, that should be studied in the future, is the decay chain $\psi(4160)\rightarrow DD_{1}\rightarrow DD^{\ast
}\pi$. In order to properly perform such a study, one should take into
account the couplings between $\psi(4160)$ and both channels $DD_{1}$ and $DD_{1}'$ (for consistency one should include not only the $D_{1}^{0}(2420)$, but also the $D_1'\equiv D_{1}^{0}(2430)$ as its pair). Moreover, a finite width for the $D_{1}^{0}(2420)$ should be considered, as well as for the very broad (although unconfirmed) resonance $D_{1}^{0}(2430)$. In this
respect, future experimental and theoretical studies along this direction are definitely needed.

\section*{Acknowledgments}
This work was supported by the \textit{Polish National Science Centre} through
the project OPUS no.~2015/17/B/ST2/01625.

\appendix

\section{\label{AA} Quadripolar cutoff}

\begin{figure}
\centering
\resizebox{!}{230pt}{\includegraphics{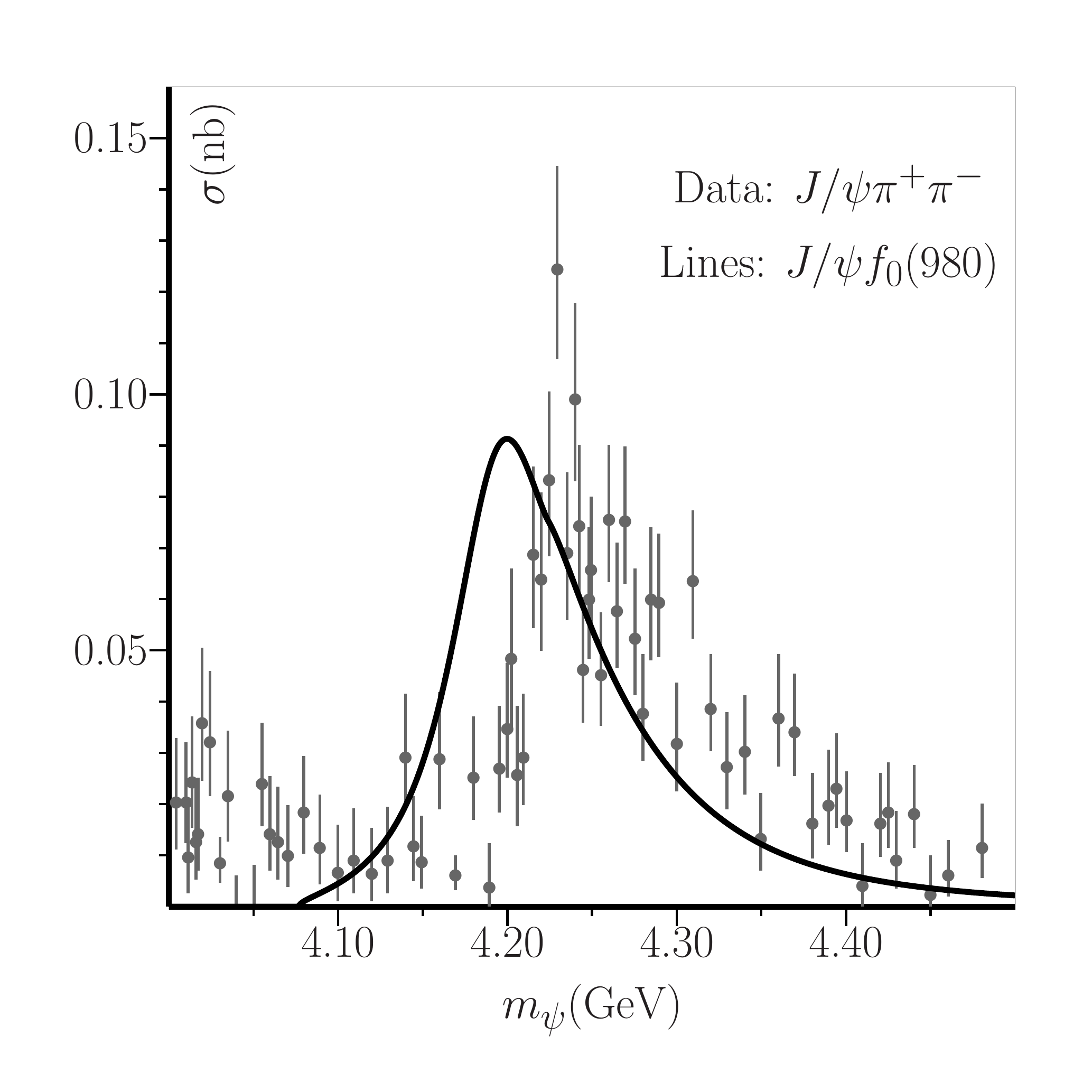}}
\caption{\label{hco} Cross-section for $e^+e^-\to\psi(4160)\to\dsds\to J/\psi f_0(980)$, using the loop-driven decay if Fig.~\ref{4vertex}, compared with the experimental cross-section $e^+e^-\to J/\psi \pi^+\pi^-$ in Ref.~\cite{prl118p092001}, for the hard cutoff case. See text for details.}
\end{figure}

In this Appendix, we study the case in which, instead of the gaussian cutoff-function in Eq.~\eqref{ff}, we use the quadripolar form given by
\be
\label{hcoff}
f_\Lambda(\vec{k}_j^2)=\Big(1+\frac{\vec{k}_j^4}{\Lambda^4}\Big)\ .
\ee 
The procedure to adjust the free parameters is as described for the gaussian case, and in the same way, the final behavior does not change qualitatively for the specific choice of $\Lambda$. As before, we choose the value $\Lambda=450$ MeV. The corresponding partial couplings and seed mass are
\be
\label{parhco1}
\begin{split}
&g_{\psi PP}\simeq 1.910\ ,\\
&g_{\psi PV}\simeq 0.881 \mathrm{\ GeV}^{-1}\ ,\\
&g_{\psi VV}\simeq 1.992\ ,\\
&m_0=4245 \mathrm{\ MeV}\ ,
\end{split}
\ee
that lead to a peak in the total spectral function with mass and width 4191 MeV and 70 MeV, respectively, simulating the $\psi(4160)$. In order to get an amplitude in the $\vpsif0$ channel similar to Fig.\ref{jppnormal}, i.e., enough small not to be seen in the data, we choose $g_{\psi J f_0}\simeq 0.0134\mathrm{\ GeV}^{-1}$, and finally, in order to compare the effect described in Sec.~\ref{inter} with data, we adjust $\lambda=1.5 \mathrm{\ GeV}^{-1}$, that is defined in Eq.~\eqref{alpha}. The parameter $\tilde{\Lambda}$, which enters in the Eq.~\eqref{hcoff} above, for channel $\vpsif0$, was varied between $450$ MeV and $10$ GeV, giving very similar results. We set it to be 1 GeV$^{-1}$. The final cross section, computed using the same equations as to Fig.~\ref{main}, is plotted in Fig.~\ref{hco}.

Figure \ref{hco} shows a shift in the peak upwards, form 4.191 GeV (corresponding to the total cross section) to about 4.2 GeV, which is not as striking as the shift seen in Fig.~\ref{main}, using the exponential cutoff-function. This might be due to the fact that the underlying pole position is now
\be
4192.4-i39.2 \mathrm{\ MeV}\ ,
\ee 
which is about 7 MeV lower than the corresponding pole in Table \ref{pvar}. We stress that, although not as significant as the shift seen in Fig.~\ref{main}, the effect of the loop-driven decay discussed in Sec.~\ref{inter} is still seen, and thus worth further studies.\\

The exponential form used in the main text emerges naturally from various
microscopic approaches. However, our results do not strongly depend on the
precise choice of the vertex function, as long as it is smooth and at the same
time falls sufficiently fast, see later on. A hard cutoff (i.e., a step
function) is not feasible, because it would imply that the spectral function
would fall abruptly to zero above a certain threshold; this unphysical
behavior does not lead to any satisfactory description of data when using our
model, see e.g.~Ref.~\cite{npa931p38}. Similarly, the avoidance of a form factor by using
an at least three-time subtraction scheme is also not a good strategy, as
previously studied in\ Ref.~\cite{npa931p38} (the determination of all 3 subtraction
constant is subject to uncertainties; in addition, the interaction at
low-energy is not expected to be local, but should reflect the finite
dimension of mesons. A smooth form factor is a useful -albeit rather simple-
way to take this feature into account).

It should be also underlined that our approach is a model of QCD, therefore
the value of $\Lambda$ is not the maximal value for momentum $k$ (in other
words, $\Lambda$ is not -strictly speaking- a high energy cutoff). When $k$ is
larger than $\Lambda$, that particular decay is suppressed as physical
consequence of the nonlocal interaction between the decaying meson and its
decay products (all of them are extended objects). The momentum $k$ can take
any value from $0$ to $\infty$, even arbitrarily larger than $\Lambda$. In
particular, the normalization of the spectral function (a very important
feature of our approach) involves an integration up to $k\rightarrow\infty$.
Of course, even if it is allowed to take $k$ arbitrarily large from a
mathematical point of view, our model is physically limited: since only a
single resonance is taken into account, we expect that it is valid up to about
$4.3$ GeV.

\section{\label{AB} On the $J/\psi \pi ^{+}\pi ^{-}$ final state of the $Y(4260)$}

The latest PDG entry for the $Y(4260)$ enhancement reports that it is ``seen'' in channel $J/\psi
f_{0}(980)\rightarrow J/\psi \pi ^{+}\pi ^{-}$, but no average or fit is given for its branching ratio. The only presented measurement is $0.17\pm 0.13$, from Ref.~\cite{prd86p051102}, but it is also stated that the systematic error for this value is lacking at present, showing that a future
experimental determination is needed.

Nevertheless, as explained in the main text, a more comprehensive study of the $J/\psi f_0\to J/\psi\pi^+\pi^-$ decays should include not only the $f_{0}(980)$, but other scalar mesons such as the $f_{0}(500),$ $f_{0}(1370),$ $f_{0}(1500),$ and $f_{0}(1710)$ as well. Given that all decay chains 
\be
\label{fdecays}
\psi (4160)\rightarrow D_{s}^{\ast }\bar{D}_{s}^{\ast }\rightarrow J/\psi
f_{0}\rightarrow J/\psi \pi ^{+}\pi ^{-}
\ee
contribute to the final spectrum, one should perform a coherent sum involving all $f_0$'s. However, the involvement of the $\dsds$ loops in the decays \eqref{fdecays}, guarantees that the final peak is expected to be close to the $\dsds$ threshold in each case.

There is another important point concerning the $J/\psi \pi ^{+}\pi ^{-}$
final state. At present, the resonant and non-resonant contributions for the 
$Y(4260)$ signal in the $J/\psi \pi ^{+}\pi ^{-}$ channel are not clearly
estimated, although it is known that they both exist. In Ref.~\cite{prd86p051102}, it is stated:

\begin{quote}
\textquotedblleft \textit{The mass distribution near }$1$\textit{\ GeV}$%
/c^{2}$\textit{\ suggests coherent addition of a nonresonant }$\pi ^{+}\pi
^{-}$\textit{\ amplitude and a resonant amplitude describing the }$%
f_{0}(980) $\textit{. If the peak near }$950$\textit{\ MeV/}$c^{2}$\textit{\
is attributed to a nonresonant amplitude with phase near }$90^{\circ }$%
\textit{, the coherent addition of the resonant }$f_{0}(980)$\textit{\
amplitude, in the context of elastic unitarity, could result in the observed
behavior, which is similar to that of the }$I=0$\textit{\ }$\pi ^{+}\pi ^{-}$%
\textit{\ elastic scattering cross section near 1 GeV (Fig. 2, p. VII.38, of
Ref. [26]). However, we have no phase information with which to support this
conjecture.}\textquotedblright
\end{quote}

One should therefore consider that the non-resonant background plays an
important role for the $Y(4260)$ structure in the $J/\psi \pi ^{+}\pi ^{-}$
channel and, in particular, the presence of the $D_{s}^{\ast }\bar{D}%
_{s}^{\ast }$ threshold could introduce the $\simeq 90^{0}$ phase required
to explain the signal, together with the $J/\psi f_{0}(980)$ contribution.
This would further support our claim, that it is not the decay to $J/\psi f_{0}(980)$
alone that generates the signal at about 4.23 GeV, but its \textquotedblleft
interference\textquotedblright\ with the threshold. In such background, the heavy scalar resonances $f_{0}(1500)$ and $f_{0}(1710)$, off-shell decays in combination with $J/\psi$, could be also included.

With relation to other resonant contributions to the $J/\psi \pi ^{+}\pi ^{-}
$ signal at about $4.23$ GeV, the PDG only refers to one more as
\textquotedblleft seen\textquotedblright , the $Z_{c}(3900)\pi $, estimated
to be a little higher than the $J/\psi f_{0}(980)$ ($\simeq 22\%$). Even if
we do not estimate the $Z_{c}(3900)\pi $ contribution in our approach, we
nevertheless think that it may be generated by a similar mechanism, but rather involving the nearby $DD_{1}$ thresholds. In the future, the analogous decay
chain 
\begin{equation}
\psi (4160)\rightarrow DD_{1}\rightarrow Z_{c}(3900)\pi \rightarrow J/\psi
\pi ^{+}\pi ^{-}
\end{equation}%
should be studied. Quite interestingly, the corresponding threshold is at about 
$4.28$ GeV, that is quite close to the peak of the $Y(4260)$, and therefore
may even contribute to the overall signal.

\section{\label{AC}Comment on the signal seen in $D^*D\pi$}

The $DD^{\ast }\pi $ channel that does not stem from the $D^{\ast }\bar{D}^{\ast }$ is an
OZI-suppressed mode for the $\psi (4160)$, which was seen in the experiment
(cf.~$\psi (4160)$ decays in PDG), although its contribution is not
quantified. We do not include it because in our approach we only include
OZI-allowed decays, making the exception for the $J/\psi f_{0}(980)$. In Ref.~\cite{prl122p102002}, the $DD^{\ast
}\pi $ distribution is a complex superposition of several enhancements, to
which the $Y(4260)$ contributes with a cross section of about 100 pb, which is comparable with the cross section of the $J/\psi \pi ^{+}\pi ^{-}$
distribution at 4.23 GeV in Ref.~\cite{prl118p092001}.

The process 
$e^{+}e^{-}\rightarrow D_{s}^{\ast }\bar{D}_{s}^{\ast }\rightarrow DD^{\ast
}\pi $ is strongly OZI suppressed and is therefore expected to be small within our picture (in fact, the quark content is
different in the initial and final states, contrarily to the case $D_{s}^{\ast }\bar{D}_{s}^{\ast }\rightarrow J/\psi f_{0}(980)$). It is however possible that the enhancement observed in the $DD^{\ast }\pi $
distribution, around 4.23 GeV, could be generated by a similar loop-effect
as the one we present in our manuscript, involving however the $%
DD_{1}(D_{1}^{\prime })$ and/or the $D^{\ast }D_{0}^{\ast }$ modes, rather
than the $D_{s}^{\ast }\bar{D}_{s}^{\ast }$. Similar $Y$ enhancements could be
produced with a similar mass, but not necessarily coincident with 4.23 GeV.
It would be crucial to know the value of the cross section to $%
DD_{1}(D_{1}^{\prime })$ and the $D^{\ast }D_{0}^{\ast }$ channels at the $%
Y(4260)$ mass. Although these thresholds are a bit further from the $\psi
(4160)$'s seed pole, they have large widths and, since they are $S$-wave
decays, their cross sections could still be sizable at lower masses, and
eventually high enough to generate loop-effects, e.g.~involving the $%
DD^{\ast }\pi $, at the $Y(4260)$ mass.

\section{\label{AD}Concerning the $\dsds$ cross section}

\begin{figure}
\centering
\resizebox{!}{230pt}{\includegraphics{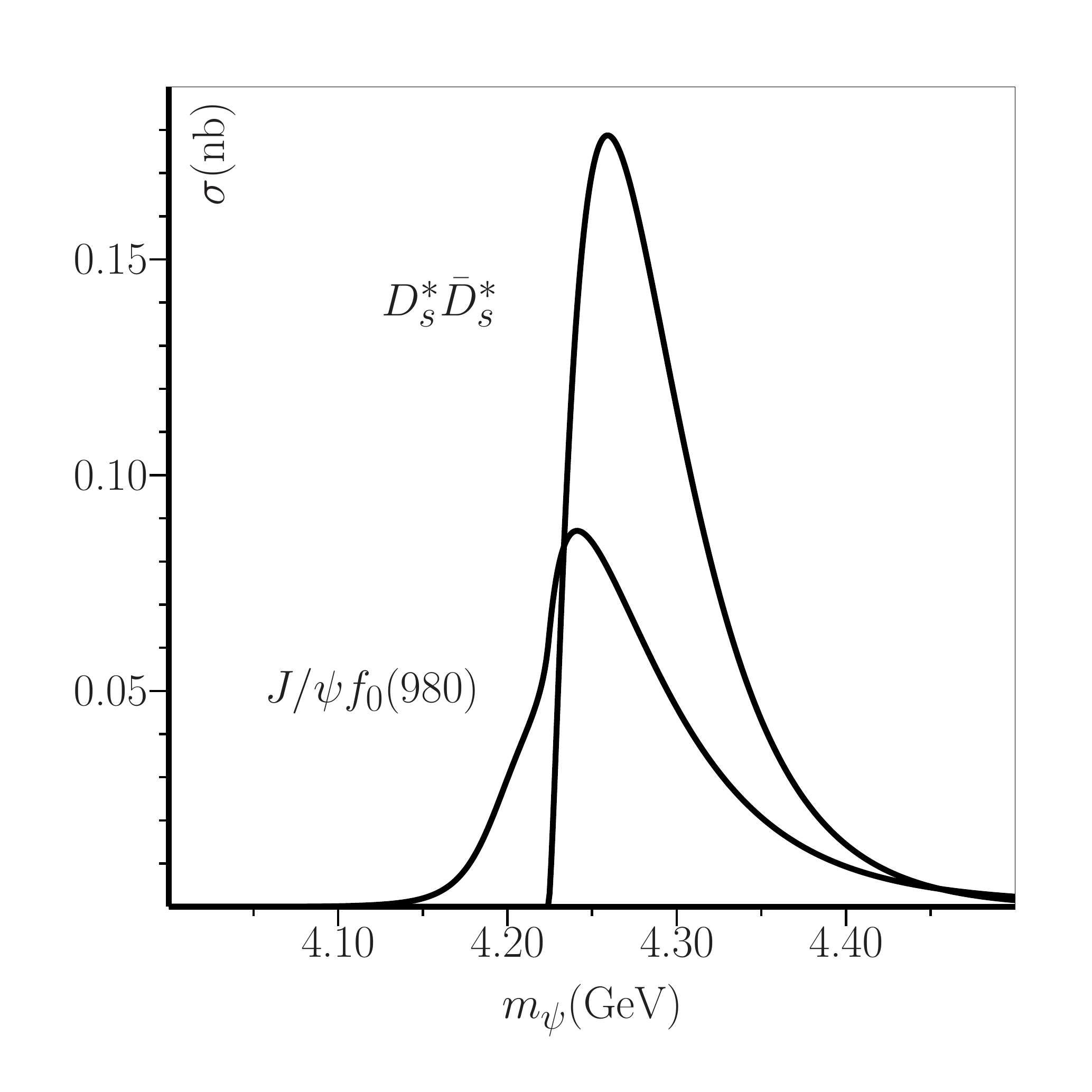}}
\caption{\label{dsasdsas} Cross-section for $e^+e^-\to\psi(4160)\to\dsds$ (higher curve), and for $e^+e^-\to\psi(4160)\to\dsds\to J/\psi f_0(980)$ as in Fig.~\ref{main} - bold line (lower curve).}
\end{figure}

Finally, we would like to discuss a delicate aspect concerning the production of the $\dsds$ pairs in our problem. Intuitively, the total production of $\dsds$ pairs has to be large enough so that a fraction of the pairs will take part in the loop-effect that leads to the $Y(4260)$. Namely, in the framework of a pertrurbative expansion, the cross section of the direct process $\psi\to\dsds$ (which is a tree-level process) is expected to be larger than $\psi\to\dsds\to J\psi f_0(980)$ (which is a one-loop process), as it
is the case within our approach, as shown in Fig.~\ref{dsasdsas}. According to our own results, the cross
section value for $\dsds$ at about 4.23 GeV (which is computed using Eq.~\eqref{cs}, using the corresponding spectral function $d_{\psi\to\dsds}$) is very close to the $\dsds\to J/\psi f_0(980)$ value, and about 4.26 GeV it is approximately the double.  In order to experimentally verify such case, by quantifying the cross section to $\dsds$, one has to assume further that most of the produced $\dsds$ pairs do not recombine into other mesons. Indeed, since the $\dsds$ is an OZI-allowed decay channel, it is natural to expect that its cross section is higher that other type of decays.

Likewise, if the final state $DD^{\ast }\pi$ should come from $DD_1$ or $DD_0^*$, via a similar mechanism, their production rate would have to be larger than for the $DD^{\ast }\pi$, and their respective cross sections expected to be higher. The $Y(4260)$ might
indeed be a composed signal which results from the superposition and
interference of different enhancements, with origin in the same $\psi (4160)$
(the only pole in the vicinity). Such phenomena are not in contradiction
with our presented ideas, but they are out of the scope of the present manuscript.

\end{document}